\documentclass[prl,aps,showpacs,superscriptaddress,twocolumn,floatfix,
nofootinbib]{revtex4}

\usepackage{amsmath,amsfonts,amssymb}
\usepackage{graphicx}

\def\3{2.8in}    
\def\2{2.5in}
\def\4{3.0in}

\def \beq {\begin{equation}}
\def \eeq {\end{equation}}

\begin{document}

\title{Dynamic Nuclear Polarization with Single Electron Spins}
\author{J. R. Petta}\altaffiliation{These authors contributed
  equally.}\affiliation{Department of Physics, Harvard University, 17
  Oxford St., Cambridge, MA 02138}\affiliation{Department of Physics,
  Princeton University, Princeton, New Jersey 08544}
\author{J. M.  Taylor}\altaffiliation{These authors contributed
  equally.}\affiliation{Department of Physics, Harvard University, 17
  Oxford St., Cambridge, MA 02138}\affiliation{Department of Physics,
  Massachusetts Institute of Technology, Cambridge, MA 02139}
\author{A. C. Johnson} \affiliation{Department of Physics, Harvard
  University, 17 Oxford St., Cambridge, MA 02138} \author{A.
  Yacoby}\affiliation{Department of Physics, Harvard University, 17
  Oxford St., Cambridge, MA 02138}
\author{M. D. Lukin}\affiliation{Department of Physics,
  Harvard University, 17 Oxford St., Cambridge, MA 02138}
\author{C. M. Marcus}\affiliation{Department of Physics, Harvard University, 17 Oxford St., Cambridge, MA 02138}
\author{M.~P.~Hanson} \author{A.~C.~Gossard} \affiliation{Materials
  Department, University of California, Santa Barbara, California
  93106}

\begin{abstract}
  We polarize nuclear spins in a GaAs double quantum dot by
  controlling two-electron spin states near the anti-crossing of the singlet (S) and $m_S$=+1 triplet ($T_+$)
  using pulsed gates. An initialized S state
  is cyclically brought into resonance with the $T_+$ state, where hyperfine fields drive rapid rotations
  between S and $T_+$, `flipping' an electron spin and `flopping' a
  nuclear spin. The resulting Overhauser field approaches 80 mT, in agreement with a simple rate-equation model.  A self-limiting pulse sequence is
  developed that allows the steady-state nuclear polarization to be set using a gate voltage.
\end{abstract}

\maketitle

Semiconductor quantum dots share many features with real atoms: they possess discrete electronic energy states,
obey  Hund's rules as the states are filled, and can be coupled to one another, creating artificial molecules
\cite{dotreview, Hanson_RMP_2007}. In contrast to atoms, where electrons are coupled to a single nucleus,
electrons within quantum dots interact with many (typically $\sim$$10^6$) lattice nuclei \cite{Hanson_RMP_2007}.
The dynamics of spin systems comprising few electrons interacting with many nuclei is an interesting and complex
many-body problem in condensed matter physics
\cite{Chen_Condmat_2007,Coish_JAP_2007,Rudner_condmat_2007,Christ_PRB_2007}.

\begin{figure}[t]
\begin{center}
\includegraphics[width=8.4cm]{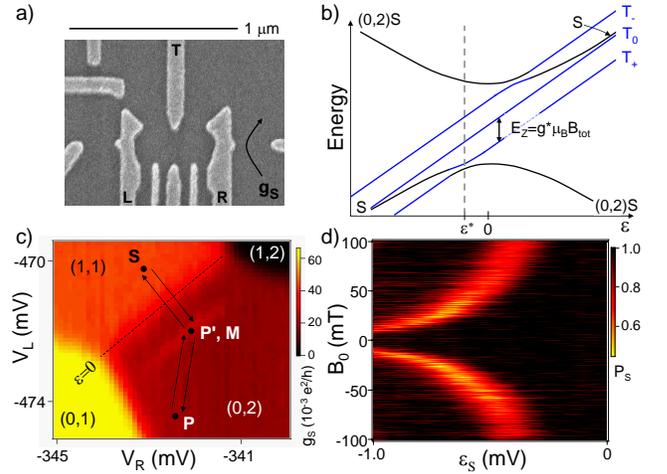}
\end{center} \vspace{-0.5 cm} \caption{(a) SEM image of a device similar to the one used in this experiment. Gates L, R set the
occupancy of the double dot while gate T tunes the interdot tunnel coupling. A QPC with conductance, $g_{\rm
S}$, senses charge on the double dot. (b) Charge sensor conductance, $g_{\rm S}$, measured as a function of
$V_L$ and $V_R$. A pulse sequence used to polarize nuclei is superimposed on the charge stability diagram (see
text). The bright signal at point M that runs parallel to the $\epsilon$=0 line is a result of Pauli-blocked
(1,1)$\rightarrow$(0,2) charge transitions and reflects $S$-$T_+$ mixing at point S. A background plane has been
subtracted from the data. (d) Singlet return probability $P_S$ for separation time $\tau_{\rm S}$=200$\rm$ ns,
as a function of separation position $\epsilon_{\rm S}$ and applied field $B_{\rm 0}$. The field dependent
``spin funnel'' corresponds to the $S$-$T_+$ anti-crossing position, $\epsilon_{\rm S}$=$\epsilon^*$.}
\end{figure}

Electrically controlled nuclear polarization in semiconductor microstructures has been investigated in
gate-defined quantum point contacts (QPCs) in GaAs, where nuclear polarization driven by scattering between
spin-polarized edge states induced hysteresis in conductance as a function of magnetic field
\cite{Wald_PRL_1994,Yusa_Nature_2005}. In few-electron quantum dots, transport in the so-called Pauli blockade
regime, which requires a spin flip for conduction, can exhibit long-time-scale oscillations and bi-stability,
also understood to result from a build-up and relaxation of nuclear polarization during spin-flip-mediated
transport \cite{Ono_Science_2002,Ono_PRL_2004,Koppens_Science_2005,Rudner_condmat_2007}. Depending on
conditions, sizable polarizations can result from transport in this regime \cite{Baugh_condmat_07051104}. Spin
relaxation at low magnetic fields, (time scale $T_1$) and spin dephasing (time scale $T_2^*$) in GaAs double
quantum dots are limited by hyperfine interactions with host nuclei
\cite{Johnson_Nature_2005,Petta_Science_2005,Koppens_Nature_2006,Erlingsson_PRB_2001,Khaetskii_PRL_2002}. It has
been suggested that electron spin dephasing times may be extended by cooling nuclear spins through nuclear
polarization and projective measurement \cite{Coish,Atac,Rudner_condmat_2007}. In addition, nuclear spins may be
used as a quantum memory due to long nuclear spin coherence times \cite{taylor}. For these reasons, it is
desirable to have the ability to control interactions between a \textit{single} confined electron spin and a
bath of quantum dot nuclear spins \cite{lukin}.

In this Letter, we use fast pulsed-gate control of two-electron spin states in a double quantum dot to create
and detect nuclear polarization.  Rather than using spin-flip-mediated transport with an applied bias
\cite{Ono_PRL_2004,Baugh_condmat_07051104}, we use a cycle of gate voltages to prepare a spin singlet ($S$)
state, induce nuclear flip-flop at the singlet-triplet ($S$-$T_+$) anti-crossing, and measure the resulting
nuclear polarization from the position of that anti-crossing. In this way, dynamic nuclear polarization (DNP) is
achieved (and measured) by the controlled flipping of single electron spins. Additionally, we demonstrate a
self-limiting pulse cycle with fixed period that allows voltage-controlled ``programming" of steady-state
nuclear polarization. Experimental results are found to be in qualitative agreement with a simple rate-equation
model.

Measurements are carried out using a gated-defined GaAs double quantum dot \cite{Petta_Science_2005}.  The
device is configured near the (1,1)-(0,2) charge transition (see Fig.\ 1(a)). A QPC charge sensor is used to
read out the time-averaged charge configuration of the dot, from which spin states can be inferred. Relevant
energies of spin states as a function of detuning, $\epsilon$, are shown in Fig.\ 1(b). The (1,1) singlet, $S$,
and (0,2) singlet, denoted $(0,2)S$, hybridize near $\epsilon$=0 due to interdot tunneling. Triplets of (0,2)
are not considered, due to the large singlet-triplet splitting, $E_{ ST}$$\sim$400 $\mu$eV. The triplet states
of (1,1), $T{_+}$, $T{_-}$, and $T{_0}$, do not hybridize with $(0,2)S$ due to spin selection rules. An external
magnetic field, $\vec{B}_0$, applied perpendicular to the plane of the two-dimensional electron gas (2DEG), adds
to the hyperfine field, $\vec{B}_{\rm nuc}$, resulting in a total effective Zeeman field $\vec{B}_{\rm
tot}$=$\vec{B}_0$+$\vec{B}_{\rm nuc}$. The Zeeman energy, $E_{\rm Z}$=$g^*\mu_{\rm B}B_{\rm tot}$, lifts the
triplet-state degeneracy, where $B_{\rm tot}$=$|\vec{B}_{\rm tot}|$ and $g^*$$\sim$-0.4 is the effective
electronic g-factor in GaAs. The nuclear Zeeman field is proportional to nuclear polarization and has a
magnitude $|\vec{B}_{\rm nuc}|$=5.2 T for fully polarized nuclei in GaAs \cite{Paget_PRB_1977}.

Hyperfine fields rapidly drive transitions between $S$ and $T_0$ on a 10 ns time scale at large negative
detuning, where these states become degenerate \cite{Koppens_Science_2005,Petta_Science_2005}. However,
transitions between $S$ and $T_0$ do not polarize nuclei since there is no change in the total spin component
along the field, $\Delta$$m_S$=0. In contrast, the transition from $S$ to $T_+$ involves a change in electron
spin, $\Delta$$m_{S}$=1. When driven by the hyperfine interaction, this electron spin `flip' is accompanied by a
nuclear spin `flop' with $\Delta$$m_{I}$=-1. The change in $B_{\rm nuc}$ associated with this flip-flop
transition is along the direction of external field, taking into account the positive nuclear g factors for Ga
and As \cite{Hanson_RMP_2007}. The cyclic repetition of the transition from $S$ to $T_+$ can thereby lead to a
finite time-averaged $B_{\rm nuc}$ oriented along the external field.

\begin{figure}[t]
\centering
\includegraphics[width=8.4cm]{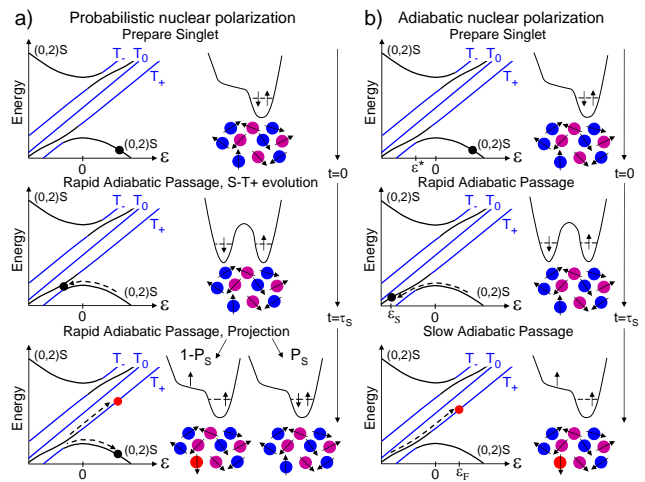}
\vspace{-0.25 cm}\caption{(a) Probabilistic nuclear polarization
  sequence. (b) Adiabatic nuclear polarization sequence (see text).}
\end{figure}

The value of detuning where the $S$-$T_+$ anti-crossing occurs, denoted $\epsilon^*$ (see Fig.~1(b)), is a
sensitive function of $B_{\rm tot}$ for $\mid$$B_{\rm tot}$$\mid$$\leq$80 mT, providing a straightforward means
of measuring $B_{\rm nuc}$ within that range. To calibrate the measurement, the dependence of $\epsilon^*$ on
external field amplitude $B_{\rm 0}$ is measured using the pulse sequence shown in Fig.~1(c): The $(0,2)S$ state
is first prepared at point P. A delocalized singlet in (1,1) is created by moving to point S (detuning
$\epsilon_{\rm S}$) via point P$'$. The system is held at point S for a time $\tau_{\rm S}$$\gg$$T_{\rm 2}^*$
then moved to point M and held there for the longest part of the cycle. The sequence is then repeated. When
$\epsilon_{\rm S}$=$\epsilon^*$, rapid mixing of  $S$ and $T_+$ states occurs.  When the system is moved to
point M, the (1,1) charge state will return to (0,2) only if the separated spins are in a singlet configuration.
The probability of being in the singlet state after time $\tau_{\rm S}$ thus appears as charge signal---the
difference between the (1,1) and (0,2) charge states---detected by measuring the time averaged QPC conductance,
$g_{\rm S}$ (see Fig.\ 1(a)). Figure~1(c) shows $g_{\rm S}$ as a function of $V_{\rm L}$ and $V_{\rm R}$ with
this pulse sequence applied. The field dependence of this signal is shown in Fig.\ 1(d), which plots the
calibrated singlet state probability, $P_S$, as a function of $B_{\rm 0}$ and $\epsilon_{\rm S}$. In Fig.\ 1(d),
$P_S$$\sim$0.7 at the $S$-$T_+$ degeneracy, corresponding to a probability of electron-nuclear flip-flop per
cycle $(1-P_S)$$\sim$0.3. The position of the anti-crossing, $\epsilon_{\rm S}$=$\epsilon^*$, becomes more
negative as $B_{\rm 0}$ decreases toward zero (Fig.~1(d)), as expected from the level diagram, Fig.~1(b).

An alternative sequence that (in principle) deterministically flips one nuclear spin per cycle and allows
greater control of the steady-state nuclear polarization is shown in Fig.\ 2(b): In this case, the initial
$(0,2)S$ is separated quickly ($\sim$1 ns), to a value of detuning beyond the $S$-$T_+$ resonance,
$\epsilon_{\rm S}$$<$$\epsilon^*$. Since the time spent at the $S$-$T_+$ resonance during the pulse is short,
the singlet is preserved with high probability. Next, detuning is brought back toward zero, to a value
$\epsilon_{\rm F}$, on a time scale slow compared to $T_2^*$ ($\sim$100 ns). This converts $S$ to $T_+$ and
flips a nuclear spin each cycle. Detuning is then rapidly moved to the point M (again, over a time $\sim$1 ns).
While slightly different, both pulse sequences rely on bringing the $S$ and $T_+$ states into resonance; the
otherwise large difference between nuclear and electron Zeeman energies would prevent flip-flop processes
\cite{Imamoglu_PRL_2003}. We now explore characteristics of the resulting nuclear polarization for each of the
processes illustrated in Fig.~2.

\begin{figure}[t]
\centering
\includegraphics[width=8.4cm]{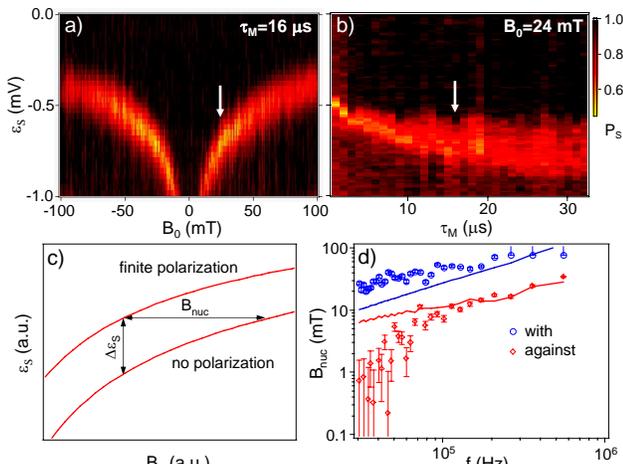}
\vspace{-0.75 cm}\caption{(a) Singlet return probability $P_S$ as a function of applied field $B_{\rm 0}$ and
detuning at the separation point,  $\epsilon_{\rm S}$. (b) Dependence of the $S$-$T_+$ anti-crossing position on
measurement duration $\tau_{\rm M}$, with $B_{\rm 0}$=24 mT. A decrease in $\tau_{\rm M}$ increases the
polarization rate, shifting the $S$-$T_+$ resonance condition to more positive $\epsilon_{\rm S}$. (c) Schematic
diagram illustrating the effect of nuclear polarization on the spin funnel. Polarization increases $B_{\rm
nuc}$, shifting the funnel to more positive $\epsilon_{\rm S}$. The value of $B_{\rm nuc}$ is extracted by
measuring the shift in the $S$-$T_+$ position relative to the position without polarization. (d) $B_{\rm nuc}$
as a function of cycle frequency, $f$, measured sweeping $\epsilon_{\rm S}$ to increasing values (``with") and
decreasing values (``against"). In the increasing sweep direction, the motion of $\epsilon_{\rm S}$ coincides
with the polarization-induced motion of the $S$-$T_+$ anti-crossing, resulting in a higher polarization. Solid
lines are predictions from a nonlinear diffusion model (see text).}
\end{figure}

We begin by examining the statistical polarization sequence shown in Fig.~2(a). We first measure $P_S$ as a
function of $B_{\rm 0}$ and $\epsilon_{\rm S}$, with $\tau_{\rm S}$=100 ns, $\tau_{\rm P}$=300~ns, $\tau_{\rm
P'}$=100~ns, $\tau_{\rm M}$=32~$\mu$s. These data, shown in Fig.~3(a), map out the $S$-$T_+$ anti-crossing in
the limit of minimal measurement induced polarization (polarization is negligible for $\tau_{\rm M}$$>$30
$\mu$s, as will be shown below). A steady-state nuclear polarization at $B_{\rm 0}$=24 mT results in a shift in
the position of $\epsilon^*$, depending on the measurement time $\tau_{\rm M}$, which determines the cycle
period (Fig. 3b). As $\tau_{\rm M}$ decreases, the position of $\epsilon^*$ moves to larger values of detuning,
indicating an increase in the average $B_{\rm nuc}$. For $\tau_{\rm M}$$>$30 $\mu$s the value of $\epsilon^*$
saturates to its unpolarized value. Values for $B_{\rm nuc}(\tau_{\rm M})$ can be extracted from calibrating the
shift in $\epsilon^*$ using the data in Fig.~3(a), as illustrated in Fig.~3(c).

We find that both the position, $\epsilon^*$, and the width of the region where $P_S(\epsilon_{\rm S})$ is
reduced below unity---marking the mean and fluctuations of the position of the $S$-$T_+$ anti-crossing---depend
on the sweep direction used for data acquisition. In the present measurements, $\epsilon_{\rm S}$ is swept on
the inner loop and $\tau_{\rm M}$ is increased on the outer loop. Sweeping $\epsilon_{\rm S}$ to more negative
values (downward in Fig.~3(b)), i.e., {\em against} the movement of $\epsilon^*$ with increasing polarization,
results in smaller nuclear polarization and a narrower resonance width compared to sweeping $\epsilon_{\rm S}$
to more positive values (not shown), i.e., {\em with} the motion of $\epsilon^*$ as polarization builds up. The
extracted $B_{\rm nuc}$ as a function of cycle frequency, $f$=$1/(\tau_{\rm P}$$+$$\tau_{\rm P'}$$+$$\tau_{\rm
S}$$+$$\tau_{\rm M}$) is shown in Fig.~3(d) for the two sweep directions of $\epsilon_{\rm S}$.  In both cases,
polarization increases with $f$ reaching a maximum value of $B_{\rm nuc}$$\sim$80 mT (see Fig.~3(d)).

Qualitative features of the statistical DNP cycle (Fig.~2(a)) are accounted for by nonlinear rate equations for
the nuclear polarization in the left and right dots, $P_{L,R}$. Within this model, a polarization-dependent
probability per cycle of electron nuclear flip-flop, $p_{\rm flip}(P)$, where $P$=$P_L$+$P_R$,  induces nuclear
polarization, while out-diffusion relaxes polarization with rates $\Gamma_L$ and $\Gamma_R$ in the left and
right dots. Nonlinearity arises from the dependence of $p_{\rm flip}$ on the distance $\Delta$=$\epsilon_{\rm
S}$-$\epsilon^*$ from the separation point to the anti-crossing, which in turn depends on $B_{\rm tot}$, and
hence $P$.  The relevant energy scale for $\Delta$ is the width of the $S$-$T_+$ anti-crossing, $\Omega$.
Writing $\Omega$=$\hbar \sqrt{2} \cos(\theta) / T_2^*$, where $\theta$=$\arctan(\frac{2 t}{\epsilon_{\rm S} -
\sqrt{4 |t|^2+(\epsilon^*)2}})$ is the adiabatic angle \cite{Taylor_PRB_2007} that accounts for charge state
mixing near $\epsilon$=0, $T_2^*$ is the inhomogeneously broadened transverse relaxation time at the $S$-$T_+$
anti-crossing, and $t$ is the interdot tunnel coupling, we find that the probability of a spin flip per cycle is
given by $p_{\rm flip} (P)$=$[\Delta^2/(2 \Omega^2) +2]^{-1}$. Note that $p_{\rm flip}$=1/2 at the
anti-crossing, $\epsilon_{\rm S}$=$\epsilon^*$. Polarizations $P_L, P_R$ of the left and right dots then evolve
according to coupled diffusion equations,
\begin{equation} \frac{dP_{L,R}(t)}{dt} = - \Gamma_{L,R} P_{L,R} + \frac{p_{\rm flip}(P)\,f \,
N_{L,R}}{3/2[N_{L}+N_{R}]^2},
\end{equation} where $N_L$ and
$N_R$ are the number of nuclei in the left and right dots. The factor of 3/2 reflects the spin-3/2 Ga and As
nuclei. Results of the model using experimental parameters $N_L$$\sim$$N_R$$\sim$3$\times$$10^6$, with
$\Gamma$=$\Gamma_L$+$\Gamma_R$ as a fit parameter (assuming $\Gamma_L$=$\Gamma_R$) are shown in Fig.\ 3(d). The
interdot tunnel coupling, $t/(g^* \mu_B)$$\sim$390 mT, is determined from measurements of the charge occupancy
as a function of detuning near the (1,1)-(0,2) interdot charge transition \cite{dicarlo}.

Qualitative agreement between experiment and model are seen in the dependence of polarization on cycle frequency
and sweep direction. The fit gives $\Gamma \sim 0.3$~s$^{-1}$ which is consistent with out-diffusion
predominantly perpendicular of the plane of the 2DEG, $\Gamma$$\sim$$D/d^2$, taking reasonable values for the
electron gas thickness $d$$\sim$6 nm and the nuclear diffusion constant $D$$\sim$$10^{-13}$ cm$^2/$s
\cite{Paget_PRB_1982}. Polarizations exceeding the $\sim$1$\%$ observed here can be achieved by increasing the
cycle rate. Baugh \textit{et al.} obtained polarizations approaching 20$\%$ in the Pauli blockade regime, which
was associated with a tunneling rate of 10$^8$ Hz, several orders of magnitude faster than the polarization
cycle frequencies employed here \cite{Baugh_condmat_07051104}.

\begin{figure}[t]
\centering
\includegraphics[width=8.4cm]{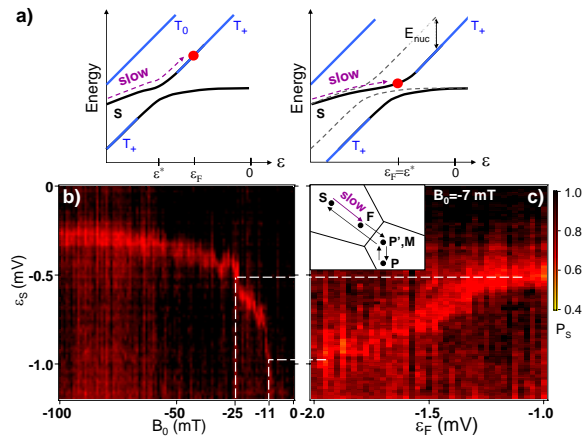}
\vspace{-0.75cm}\caption{(a) Self-limiting adiabatic polarization sequence. Left: Starting with $B_{\rm nuc}$=0,
the sequence polarizes nuclear spins and $B_{\rm nuc}$ increases, shifting the $S$-$T_+$ resonance to more
positive $\epsilon$. Right: Steady state is reached when $\epsilon^*$=$\epsilon_{\rm F}$. Dashed grey lines show
the energy level configuration before nuclear polarization. (b) Calibration spin funnel acquired using the pulse
sequence (Fig.\ 1(c)). (c) $P_S$ as a function of $\epsilon_{\rm F}$ measured with the adiabatic polarization
sequence applied. The complete pulse sequence consists of three adiabatic polarization cycles (a single
adiabatic polarization cycle is shown in the inset) followed by one measurement cycle (shown in Fig.\ 1(c)). The
combined pulse sequence length is $\sim 6.4$ $\mu$s. Steady state hyperfine field $B_{\rm nuc}$ changes from 4
mT to 18 mT as set-point $\epsilon_{\rm F} $ is moved from $-2$ mV to $-1.3$ mV.}
\end{figure}

We next investigate a self-limiting adiabatic DNP cycle that allows steady-state nuclear polarization to be set
by a gate voltage, as illustrated in Fig.~2(b). The cycle is shown in terms of motion along energy levels in
Fig.\ 4(a) and motion within the charge stability diagram in the inset of Fig.~4(c).  The new feature in this
sequence, compared to the statistical DNP cycle (Fig.~2(a)), is that the cycle passes {\em through} the
anti-crossing at $\epsilon^*$ quickly during the pulse to the separation point $\epsilon_{\rm S}$, then slowly
back to a point $\epsilon_{\rm F}$, before moving quickly to the measurement point, $\epsilon_{\rm M}$. The
cycle then continues through $\epsilon_{\rm P}$ and $\epsilon_{\rm P'}$ as before. The position of
$\epsilon_{\rm F}$ is chosen so that at low polarization it is to the right of $\epsilon^*$, as seen in the left
diagram in Fig.~4(a). Adiabatic crossing of the $S$-$T_+$ anti-crossing in the return direction results in a
single electron-nuclear flip-flop and leads to nuclear polarization upon cycling. As polarization builds,
$\epsilon^*$ moves to more positive values of detuning, until it coincides with the set-point, $\epsilon_{\rm
F}$, as illustrated on the right panel of Fig.\ 4(a). At that point, the build-up of polarization stops.

To both induce and measure a steady-state nuclear polarization, a sequence of three DNP cycles (Fig.~4(c),
inset) followed by one measurement cycle (Fig.~1(c)) is iterated. We first calibrate changes in the position
along $\epsilon$ with changes in $B_{\rm nuc}$ by measuring the singlet return probability $P_S$ as a function
of $\epsilon_{\rm S}$ and applied field $B_0$, showing how $\epsilon^*$ depends on $B_{\rm tot}$ in the absence
of nuclear polarization (Fig.~4(b)). Figure 4(c) shows that for the specific case of $B_0$=7 mT, the
steady-state nuclear polarization can be controlled by the position of the set-point $\epsilon_{\rm F}$, ranging
from $B_{\rm nuc}$$\sim$4 mT for $\epsilon_{\rm F}$=-2 mV to $B_{\rm nuc}$$\sim$18 mT for $\epsilon_{\rm
F}$=-1.3 mV. For $\epsilon_{\rm F}$$>$-1.3 mV, $B_{\rm nuc}$ saturates at $\sim$18 mT. In these measurements,
the low fields involved were necessary to allow calibration using plots like Fig.~4(a), which flatten and become
insensitive to changes in $B_{\rm nuc}$ at higher fields. This can be overcome by using single-electron spin
resonance \cite{Koppens_Nature_2006, EDSR}, which allows calibration of $B_{\rm nuc}$ at arbitrary fields.
Polarization with the adiabatic pulse sequence is less than expected given the cycle rate (it should be twice as
efficient as the probabilistic pulse sequence at a given cycle frequency, $f$) and did not grow when the cycle
frequency was increased beyond the value used to acquire the data in Fig.\ 4(c). Further experiments are
required to determine if dark state formation is limiting polarizations obtained with this pulse sequence, where
successive interactions at the S-$T_+$ degeneracy take place on $\sim$300 ns time scales, which are much shorter
than nuclear spin decoherence times \cite{Imamoglu_PRL_2003}. 

This work was supported by DARPA, DTO, NSF-NIRT (EIA-0210736), Harvard Center for Nanoscale Systems, and Princeton University. Research at UCSB supported in part by QuEST, an NSF Center.

\end{document}